\documentstyle[12pt,aasms]{article}
\begin{document}
\centerline{\bf Intergalactic UV Background Radiation Field}
\bigskip
\bigskip
\centerline{Snigdha Das and Pushpa Khare}
\centerline{Physics Department, Utkal University}
\centerline{Bhubaneswar, 751004}
\bigskip
\bigskip
\centerline{Running Title: Intergalactic UV Background Radiation Field}
\newpage
\centerline{Abstract}

We have performed proximity effect analysis of low and high resolution
data, considering detailed frequency and redshift dependence of the AGN
spectra processed through galactic and intergalactic material. We show
that such a background flux, calculated using the observed distribution
of AGNs, falls short of the value required by the proximity effect
analysis by a factor of $\ge$ 2.7. We have studied the uncertainty in
the value of the required flux due to its dependence on the resolution,
description of column density distribution, systemic redshifts of QSOs
etc. We conclude that in view of these uncertainties the proximity
effect is consistent with the background contributed by the observed
AGNs and that the hypothesized presence of an additional, dust extinct,
population of AGNs may not be necessary.


\section {Introduction}
   
   In recent years quasar absorption lines have yielded unique 
information about the physical conditions at high redshifts. The
proximity effect, which is the decrease in the number of Ly $\alpha$
forest lines having neutral hydrogen column density above a certain
minimum value, per unit redshift interval, near the QSO, has been used
to determine the intensity of the intergalactic ultraviolet background
radiation (IGUVBR) at high redshifts (Bajtlik, Duncan \& Ostriker,
1988). With a large sample of QSOs observed at intermediate resolution
Bechtold (1994) confirmed the presence of proximity effect at a high
significance level. Bechtold (1994, 1995) also considered several
sources of uncertainty in the value of the flux obtained from the
analysis of the proximity effect. Espey (1993) considered the
possibility of a higher systemic redshifts of QSOs, while Loeb \&
Eisenstein (1995) considered the possibility of quasars residing in
clusters of galaxies.  These two possibilities were also considered by
Srianand \& Khare (1996, hereafter SK96) for a large, homogeneous
sample. In addition SK96 showed that the study of proximity effect in a
sample of QSOs having damped Ly $\alpha$ absorbers along their lines of
sight provides an indirect proof of the presence of dust in such
absorbers.  All these studies used intermediate resolution data.
Proximity effect calculations using such data suffer from curve of
growth effects and the assumptions of the I model (Bajtlik et al 1988)
used in these calculations may not be strictly valid. Also line
blending is inherent in the low resolution data and therefore the column
density distribution implied by the equivalent width distribution may
be considerably different from the actual distribution (SK96). It is
therefore worthwhile exploring these effects using high resolution
data.

   Bechtold (1994) obtained the value of the intensity of the IGUVBR at the
Lyman limit, J$_{\rm \nu_{LL}}$, to be 3 J$_{21}$ (J$_{21}$ =10$^{-21}$
ergs cm$^{-2}$ s$^{-1}$ Hz$^{-1}$ sr$^{-1}$ ), assuming J$_{\rm \nu_{LL}}$
to be independent of redshift. From an analysis of high resolution data
Giallongo et al (1996) and Cooke, Espey \& Carswell (1996) found no
evidence for the redshift dependence of IGUVBR over the redshift range
of 1.7 to 4.5 and obtained J$_{\rm \nu_{LL}} \simeq  0.5\pm 0.1$ J$_{21}$
and 1$_{-0.3}^{+0.5}$ J$_{21}$ respectively. Values of J$_{\rm \nu_{LL}}$
obtained by Bechtold (1994) and Cooke et al (1996) are considerably
higher than the value expected from the distribution of visible QSOs.
It has been suggested (Fall \& Pei 1993) that the actual number of
QSOs may be larger than their observed number and that several QSOs may
be rendered invisible due to dust extinction in the intervening
absorbers.  It is also possible that the IGUVBR gets a significant
contribution from star forming galaxies (Madau \& Shull 1996, Giroux
\& Shapiro 1996). The shape of the IGUVBR in almost all the studies of
proximity effect has been assumed to be a power law having the same
slope as the UV spectra of the QSOs. This assumption is, however, not
valid due to the absorption and re-radiation of the UV photons by
galaxies and intergalactic material. Also if the IGUVBR gets
significant contribution from stellar sources then also its shape is
likely to be considerably different from a power law.

   In this paper we first study, using a large sample, of QSOs observed
at intermediate resolution as well as a sample of QSOs observed at high
resolution (section 2), the effect of assuming a more realistic shape
and redshift dependence of the IGUVBR on the value of J$_{\rm
\nu_{LL}}$  obtained from the proximity effect analysis (section 3). We
then study (section 4) the uncertainties in the value of J$_{\rm
\nu_{LL}}$ due to various possibilities mentioned above using the
sample of Ly $\alpha$ lines with measured column densities in the
spectra of QSOs observed at high resolution.

\section{Data sample}

   Our low resolution sample (LRS) is same as that used by SK96 for
proximity effect analysis. It consists of 54 QSOs observed at a
resolution between 60 to 100 km s$^{-1}$. The minimum equivalent width
limit used for this sample is 0.3 $\AA$ which is above the completeness
limit for the sample. The high resolution sample (HRS) consists of
lines observed towards 9 QSOs.  The details of the sample are given in
Table 1, which lists the emission redshift, z$_{\rm em}$, corrected
emission redshift, z$_{\rm em}^{\rm c}$, minimum observed redshift,
z$_{\rm min}$, maximum observed redshift, z$_{\rm max}$, quasar flux
f$_{\nu}$ and references for all the 9 QSOs. z$_{\rm em}^{\rm c}$ are
the average values of corrected redshifts of all available emission
lines, calculated as described by Tytler \& Fan (1992) to obtain the
systemic redshifts. z$_{\rm min}$ is the larger of the observed minimum
and the redshift corresponding to the Ly $\beta$ emission. f$_{\nu}$ is
the QSO continuum flux at the Lyman limit, in units of microjanskies.
The values are calculated by extrapolating the continuum flux at the
rest wavelength of $\lambda$(1450) to the Lyman limit. The minimum
neutral hydrogen column density cutoff for the sample is taken to be
10$^{13}$ cm$^{-2}$.

\section{Shape of IGUVBR}

   Shape of the IGUVBR due to AGNs and young galaxies is affected
considerably by the absorption by galaxies and intergalactic matter
(Bechtold et al 1987; Miralda-Escude \& Ostriker 1990).  Recently
Haardt \& Madau (1996, hereafter HM96) have shown that radiation from
re-combination within the clumpy intergalactic gas contributes
significantly to the IGUVBR. They have determined the spectrum of
IGUVBR due to AGNs at several redshifts taking into account the
absorption as well as re-radiation due to intervening material. The
ionization rate of H I due to this background is roughly 1.5 times the
rate if the recombination radiation is omitted.

   We have used the shape and redshift dependence of the IGUVBR of HM96 to
calculate the expected number of Ly $\alpha$ lines near the QSOs,
having equivalent width greater than 0.3 $\AA$ for LRS and having
column density greater than 10$^{13}$ cm$^{-2}$ for HRS. This
calculation is similar to the I model calculation of Bajtlik et al
(1988) except that we explicitly calculate the ionization rate of
neutral hydrogen at different distances from each of the QSOs in the
sample using the shape and intensity of the IGUVBR at that redshift
obtained by interpolating between the spectra given by HM96 in their
Fig. 5. The expected number of Ly $\alpha$ lines per unit redshift
interval, having neutral hydrogen column density above N$_{\rm
H\;I}^{\rm min}$, at a redshift z in the spectra of a QSO having
emission redshift z$_{e}$ is then given by
$${{\rm dN\over{dz}}}\;={\rm N_0}\rm(1+z)^\gamma \; {\rm
R^{1-\beta}}$$
 where $\gamma$ and $\beta$ describe the distribution of
Ly $\alpha$ lines away from the QSOs w.r.t.  redshift and column
density respectively, the number of lines per unit redshift interval
per unit column density interval being proportional to
(1+z)$^{\gamma}$N$^{-\beta}_{\rm{H\;I}}$. N$_0$  is the value of dN/dz
at z=0 and R is given by
 $$ {\rm{R}}\;=\;{\rm{\int\limits_{\nu_{LL}}^{\infty}{\sigma_{\nu}\;
(J_\nu(z)+{f_\nu(z_{e},z)\over{4\;\pi}})\;
dz}}\over{\rm{\int\limits_{\nu_{LL}}^{\infty} {\sigma_{\nu}\;
J_\nu(z)\;dz}}}} ,$$
being the ratio of neutral hydrogen column density in a given cloud at
redshift z if it is ionized both by QSO radiation and IGUVBR to the
column density if it is ionized by IGUVBR alone. This factor replaces
(1+$\omega$) factor used in the earlier analysis of I model, where
$\omega$ is defined as
$$ {\omega} = {\rm {f_\nu(z_{e},z)\over{4\;\pi \;J_\nu(z)}}} ,$$
Here $\sigma_{\nu}$ is the ionization crosssection of H I and
f$_\nu$(z$_{e}$,z) is the flux from QSO at the redshift z. Values of
N$_0$ (6.73 for LRS \& 23.8997 for HRS), $\gamma$ (1.810 for LRS \&
1.903 for HRS) and $\beta$ (1.5453 for HRS) for the sample are obtained
by performing a maximum likelihood analysis of the sample of lines at
distances larger than 8 Mpc from the respective QSOs, which is presumed
to be free of the effects of ionization by the QSO flux.  The total
number of expected lines in the spectra of all QSOs in the sample as a
function of relative velocity w.r.t. the QSOs is shown in Fig.1 for
HRS. The figure also includes the histogramme for the observed number
of lines with different relative velocities.

   As seen from the figure, the expected numbers of lines in the region
close to the QSO are much smaller than the observed values. This is
because the background flux is small and QSO flux is much stronger than
the background thereby making R large. The $\chi^2$  probability that
the observed number of lines with relative velocity w.r.t. the QSOs,
smaller than 12000 km s$^{-1}$ are consistent with the expected values
is $\sim$10$^{-6}$ for LRS \& 10$^{-4}$ for HRS.  As mentioned above,
higher background flux can be obtained by assuming either that large
number of QSOs are obscured due to dust extinction in intervening
absorbers or that the flux from galaxies contributes significantly to
the background. In the first case we can uniformly scale up the flux of
HM96 keeping the redshift and frequency dependence same and keeping in
mind the possibility that the true redshift and luminosity distribution
of QSOs may be different from the observed distribution and the actual
redshift and frequency dependence of the IGUVBR may be different from
that of HM96.  Good fit between the expected and observed distribution
( $\chi^2$  probability = 0.236 for LRS \& 0.822 for HRS) is obtained
for a scaling up factor $\sim$  6.3$^{+6.4}_{-2.3}$ for LRS and
5.0$^{+8.0}_{-2.3}$ for HRS. Errors are 1$\sigma$ values assuming a
Gaussian $\chi^2$ probability distribution and give the range of values
for which the $\chi^2$ probability is $\ge$ 1/$\sqrt{\rm e}$ of its
maximum value. The expected distribution is shown in Fig.1. The IGUVBR
of HM96 thus falls short of the value required by the proximity effect
by a factor of at least 2.7. 

We have explored the possibility that additional flux may be
contributed by galaxies. Madau \& Shull (1996) have estimated that at z
$\sim$ 3, galaxies which may be responsible for the generation of
metals seen in Ly $\alpha$ clouds at that redshift, can contribute a
flux of J$_{\rm \nu_{LL}}$ $\sim$ 0.5 J$_{-21}$ to the IGUVBR provided
the escape fraction of Lyman continuum photons from the galaxies is
$\ge$ 0.25. It thus seems unlikely that the background flux due to
galaxies will be sufficient to explain the proximity effect. It is
however, possible that as the Lyman alpha clouds are possibly
associated with galaxies (Lanzetta et al 1995; Boksenberg 1995) the
radiation from local stellar sources contributes significantly to the
radiation incident on the clouds. We have explored this possibility and
have calculated the expected distribution for the case when radiation
from local stellar sources contributes to the flux incident on the
clouds. Steidel (1995) from his study of a large sample of galaxies
associated with QSO absorption lines of heavy elements at z$\le$ 1
finds these galaxies to be normal in the sense of their star formation
rates. Recently Steidel et al (1996) have found a
substantial population of normal star forming galaxies at redshifts
$>$3. We have therefore taken the shape of the local radiation field
to be that given by Bruzual (1983) and assumed it to be independent of
the redshift.  We added this galactic flux to the background of HM96
and varied the absolute value of the galactic flux at 1 Ryd. The best
fit was obtained for J$_{\rm \nu_{LL}}$ (galaxy)=1.9$^{+2.4}_{-0.9}$
J$_{-21}$ for LRS and 1.5$_{-0.9}^{+2.7}$ J$_{-21}$ for HRS. The best
fit is also shown in Fig.1. These values are very large and can be
achieved only if the clouds lie at distances $<<$ 90 kpc of the
galactic centre (Giroux \& Shull 1997). The observed distances of the
clouds are almost an order of magnitude larger than this value
(Lanzetta et al 1995). We thus conclude that the HM96 spectra falls
short of the proximity effect estimates by a factor of $\ge$ 2.7 and
the additional flux needed is unlikely to be contributed by 
galaxies. Proximity effect calculations, assuming a pure power law
IGUVBR, leads to J$_{\rm \nu_{LL}}$ $\sim$ 2.5 J$_{-21}$ for LRS and
2.0 J$_{-21}$ for HRS. The expected number of lines for this case are
also shown in Fig.1. Same values are obtained for pure galactic
spectra, and are therefore highly insensitive to the detailed shape of
the flux.

\section{ Column density distribution }

   The I model used by Bajtlik et al (1988) assumes a single power law 
distribution for the neutral hydrogen column density. Bechtold (1994) 
pointed out the dependence of the derived value of the flux on the 
value of $\beta$, the required value of flux 
decreasing with decrease in $\beta$. Chernomordik \& Ozernoy (1993) 
showed that the observed equivalent width distribution can be 
explained from an assumed power law distribution of column density 
only if the power law index is 1.4 instead of the observed value. SK96 
argued that as the lines are often blended in low resolution data, an 
effective column density distribution (of blended lines) describing 
the equivalent width distribution should be used in I model 
calculations for low resolution data. High resolution data have
revealed a paucity of high column density lines and it seems likely that
the column density distribution is described by a double power law
(Petitjean et al 1993, Khare et al 1997). The double power law may,
however, be a result of the incompleteness of the sample at low column
density end caused by the loss of such lines due to blending. This has
been shown to be the case through the analysis of simulated spectra (Hu
et al 1995, Lu et al 1997), the real redshift distribution being a single
power law of index $\simeq$ -1.5 (however, see Giallongo et al 1996). As
the observed distribution is a double power law, it should be used in
the proximity effect calculations rather than a single power law.
Giallongo et al (1996) using the observed double power law obtained a
value of J$_{\rm \nu_{LL}}$ $\simeq$ 0.6 J$_{21}$ for their high
resolution sample, which further reduced to 0.5 J$_{21}$ when the
blending effect was accounted for. Double power law fit to our sample
of lines farther than 8 Mpc from the QSOs is given by

 f$_{\rm N_{ H\;I}}$ dN$_{\rm H\;I}$ $\propto$ N$_{\rm
 H\;I}^{-\beta_1}$ for N$_{\rm H\;I} < {\rm N_{\rm b}}$

\hskip 0.8 true in$\propto$ N$_{\rm H\;I}^{-\beta_2}$ for N$_{\rm H\;I}
> {\rm N_{b}}$\\ with $\beta_{1}\;=$ 0.936, $\beta_{2}\;=$ 2.1727  and
N$_{\rm b}\;=\;9.54\times10^{13}$ cm$^{-2},$ the distribution being
continuous at N$_{\rm b}$.  Near the QSOs the column density
distribution retains its shape except that the value of the column
density at the break changes with distance from the QSO as N$_{\rm
b}^{\rm near}$(z) = N$_{\rm b}\;(1+\omega)^{-1}$. The expected number
of lines within a given column density range, per unit redshift
interval, at a given redshift (near the QSO) can be obtained by
integrating the distribution given in the above equation  w.r.t. the
column density, using appropriate values of N$_{\rm b}^{\rm near}$(z).
The best fit for
HM96 is obtained for a scaling up factor of 2.0$^{+1.3}_{-0.5}$. The
distribution is shown in Fig 1. The fit is not as good as that with a
single power law, the $\chi^2$ probability being 0.197. The best fit
for pure galaxy spectra and power law is obtained for J$_{\rm
\nu_{LL}}$ $\sim$ 0.8$^{+0.3}_{-0.3}$ J$_{-21}$ and 0.8$^{+0.3}_{-0.2}$
J$_{-21}$, the $\chi^2$ probability being 0.266  and 0.267
respectively. The required value of galactic flux is, within the
allowed range, consistent with that expected from the starburst
galaxies. We therefore conclude that the HM96 spectra falls short of
the proximity effect requirements by a factor of $\ge$ 1.5. The
required extra flux may possibly be contributed by starburst galaxies.

In the following section we study the uncertainties in the background
flux calculations as a result of various factors mentioned in the
introduction. As we are interested in estimating the relative change in
the background flux we assume a power law background with slope = -1.5,
assume single power law column density distribution and use only the
HRS for the analysis.

\section{ Sources of Uncertainty in the value of J$_{\rm \nu_{LL}}$ }
 
\subsection{ Resolution}

Cooke et al (1996) have argued that line blending in general makes
detection of lines less likely, however, as the lines near the QSOs are
sparse detection is easier and effect is to increase the number of
lines near the QSOs. This effect will, however, be countered by the
increase in blending near the QSOs due to the fact that the number of
lines per unit redshift interval increases with z and therefore the
intrinsic line density near the QSOs is higher than that away from it.
One way to judge the effect of blending is to compare the results
obtained from observations with different resolutions. As noted before,
comparison with results of low resolution sample is not appropriate as
these samples (with measured equivalent widths rather than column
densities) may suffer from curve of growth effects and due to the
effective column density distribution being different than that
observed for the HRS (SK96).  Note that using $\beta$ = 1.4 for LRS
reduces the value of J$_{\rm \nu_{LL}}$ by a factor of 2, which is
larger than the difference between the values of J$_{\rm \nu_{LL}}$
obtained from the proximity effect analysis of the HRS and LRS. The
values for LRS are higher by a factor of $\sim$ 1.25. It is, therefore,
more appropriate to compare results of analysis of column density
measured samples observed at different resolutions. Our data has two
QSOs, Q1100 - 264 and Q2206 - 199 observed with very high resolution
$\le$ 8 km s$^{-1}$, while the rest of the QSOs have a resolution of
between 14 and 35 km s$^{-1}$. We have performed the analysis for the
sample excluding the lines observed towards Q1100-264 and Q2206-199
which yields J$_{\rm \nu_{LL}}$ =2.5 J$_{-21}$ which is 25$\%$ higher
than the value for the whole sample. The value of J$_{\rm \nu_{LL}}$ is
thus likely to be overestimated due to line blending. 

Cooke et al (1996) have estimated the effect of blending on the
estimated value of J$_{\rm \nu_{LL}}$ by performing proximity effect
calculations for two different values of N$_{\rm H\;I}^{\rm min}$,
differing by $\Delta$ log (N$_{\rm H\;I})= $0.5. They find little
change in the lowest reasonable flux though the best fit value of
J$_{\rm \nu_{LL}}$ increases with increase in N$_{\rm H\;I}$, specially
for z $<$ 3.5, by up to 2 orders of magnitude. Based on the lowest
reasonable flux they conclude that the change in J$_{\rm \nu_{LL}}$
values due to the change in completeness limits (N$_{\rm H\;I}^{\rm
min}$) and therefore due to line blending is less than 0.1 dex, the
flux being underestimated due to blending. It is, however, not very
clear if the difference between the two J$_{\rm \nu_{LL}}$ values is
due to the effect of blending alone.  The $\gamma$ value increases with
increase in N$_{\rm H\;I}^{\rm min}$ (Acharya and Khare 1993; Cooke et
al 1996) which means relatively more lines near the QSO for  the sample
with higher value of N$_{\rm H\;I}^{\rm min}$ which may overestimate
J$_{\rm \nu_{LL}}$ value for that sample (Cooke et al 1996). Also as
pointed out by Cooke et al (1996) taking a sample of stronger (more
saturated) lines may overestimate the effect of QSO flux as the strong
lines are relatively less sensitive to the flux. It is therefore not
very clear if the J$_{\rm \nu_{LL}}$ value for the sample with
increased completeness limit is the value for lower blending. The
effect of blending found here is stronger and in an opposite sense. We
have estimated the effect by a direct comparison of flux values
obtained by including and excluding QSOs observed with a resolution
which is considerably higher than that for the rest of the QSOs. 
 We feel that our approach
may give a direct estimate of the effect of resolution and therefore
blending. Our conclusions are based on best fit values and are at lower
redshifts. The two QSOs observed with higher resolution are at
redshifts of 2.15 and 2.55 while the average redshift of the rest of 
the QSOs is 3.07. Thus part of the difference between the flux values 
obtained for the two samples may be contributed by the redshift dependence of 
J$_{\rm \nu_{LL}}$ and it may be necessary to perform a
more detailed study on a larger sample in order to understand the effect
of blending.

\subsection{Dust in damped Ly $\alpha$ systems}

   Presence of dust in damped Ly $\alpha$ systems has been indicated by
the redder colours of QSOs having these systems along their line of
sight (Fall, Pei \& McMahan 1989; Pei, Fall \& Bechtold 1991). Pettini
et al (1994) have independently confirmed the presence of dust in these
systems through the measurement of abundance of the refractory element
Cr which appears to be depleted compared to its solar abundance. SK96
obtained yet another independent proof for the existence of dust in the
damped Ly $\alpha$ systems. They argued that the observed flux of the
QSOs having such absorbers in their lines of sight must be smaller than
the actual value as a result of which the IGUVBR flux obtained from the
proximity effect analysis of a sample of these QSOs should be lower
than that obtained from the whole sample. They confirmed this with
their sample of 54 QSOs, 16 of which had damped Ly $\alpha$ lines in
their spectra. 5 QSOs in our sample have damped Ly $\alpha$ systems
along their lines of sight. Proximity effect analysis for these yields
J$_{\rm \nu_{LL}}\simeq 1.5^{+3.3}_{-0.8}$ J$_{-21}$ which is only
marginally smaller than the value of 2.0 $^{+2.64}_{-1.01}$ J$_{-21}$
for the entire sample. The decrease in the value of J$_{\rm \nu_{LL}}$
is much smaller than that found by SK96 and may be due to the fact that
our sample is much smaller and the QSOs with damped Ly $\alpha$ systems
form more than half of the sample. Large samples will be needed to
verify the presence of and estimate the amount of dust in these
systems.

\subsection{Peculiar velocities of Quasars and/or Ly $\alpha$ clouds}

   For several QSOs, some of the lines observed on the long wavelength
side of the Ly $\alpha$ emission line can not be identified as heavy
element lines . It is possible that these are Ly $\alpha$ forest lines
with a redshift larger than the emission redshift of the QSO. The
higher redshift of the Ly $\alpha$ forest line can occur due to either
the QSO having a peculiar velocity due to its presence in a cluster
and/or the Ly $\alpha$ forest clouds infalling towards the QSO or the
cluster (Loeb \& Eisenstein 1995) or having peculiar velocities
(SK96).  The last possibility is rendered viable by the observed
clustering of Ly $\alpha$ forest clouds on velocity scales of $\le$300
km s$^{-1}$ (Srianand \& Khare 1994, Chernomordik 1995) and is also
expected if Ly $\alpha$ clouds are associated with galaxies or clusters
of galaxies as mentioned above. The modification in the expected number
of lines near the QSOs taking into account some of these effects was
evaluated by Loeb \& Eisenstein (1995) and SK96.  Here we follow the
approach of SK96 and assume that the Ly $\alpha$ clouds have a Gaussian
peculiar velocity distribution with a velocity dispersion v$_{\rm d}$.
The result will also be valid for the case of the QSO having a peculiar
velocity instead of the Ly $\alpha$ clouds. Good fit between the
observed and expected values is obtained only for v$_{\rm d}$ $>$1000
km s$^{-1}$. The best fit values of J$_{\rm \nu_{LL}}$ for v$_{\rm d}$
= 1500 \& 2000 km s$^{-1}$ are 2.5 J$_{-21}$ \& J$_{-21}$
respectively.  These velocities are too large to be due to peculiar
velocities of Ly $\alpha$ clouds and could only reflect the peculiar
velocities of QSOs.  However, such high velocities, even for QSOs, can
not be obtained for realistic values of cluster masses containing QSOs
( Loeb \& Eisenstein 1995). It thus appears that the absorption lines
with redshift larger than the emission redshifts may not be caused by
the peculiar velocities of Ly $\alpha$ clouds and/or QSOs.

\subsection{Higher systemic QSO redshifts}

   Following Espey (1993) and SK96 we also considered the possibility
that the systemic redshifts of QSOs are higher than the values used
here (Table 1). Note that we have actually used the emission redshifts
corrected for the difference in redshifts of lines of the low and high
ions, as per the prescription of Tytler \& Fan (1992). The dependence
of J$_{\rm \nu_{LL}}$ on the shift in systemic redshifts (assumed to be
same for all QSOs in the sample) is shown in Fig.2. A shift by 250 km/s
will reduce the necessary value of J$_{\rm \nu_{LL}}$ by a factor
$\simeq$ 1.4, which is roughly the discrepancy between the flux of HM96
and that required by the proximity effect.

\section{Conclusions}

We have performed the proximity effect calculations for low resolution
as well as high resolution data assuming different shapes and redshift
dependence of the IGUVBR. We find that the required intensity of the
background flux is highly sensitive to the shape of the column density
distribution used in the analysis. The use of a double power law
reduces the intensity by a factor of 2.2 from the value obtained by
using a single power law distribution. It is therefore important to
have a large sample of lines observed at high resolution in order to
accurately determine the column density distribution. Higher systemic
redshifts of the QSOs by only $\sim$ 250 km s$^{-1}$ reduce the
required intensity by a factor of 1.4. The presence of dust in damped
Lyman $\alpha$ systems on the other hand may be responsible for an
underestimate by more than 25$\%$ of the required value of the flux. A
similar effect may also be present due to the
limitation in resolution used for observing the QSOs. Pure AGN
background, processed through galaxies and intergalactic matter falls
short of the proximity effect requirements by a factor of $\ge$ 1.5.
However, considering the uncertainties in the required intensity due to
its dependence on several other factors mentioned above, this may not
be a serious discrepancy. The required value of the flux is highly
insensitive to the shape of the background.
In view of these uncertainties the proximity
effect may also be entirely accounted for by the 
radiation from the galaxies responsible for producing heavy elements 
observed in the Lyman alpha clouds.
Note that we have not taken into
account the additional uncertainties in the value of J$_{\rm \nu_{LL}}$
due to the uncertainties in the values of $\gamma$, $\beta$, QSO flux
etc (Cooke et al 96). Thus we conclude that at present there is no 
compulsive evidence
from proximity effect for a larger, dust extinct, QSO population or a
substantial contribution from galactic sources and pure AGN flux may be
adequate to explain the proximity effect.
\centerline{\bf Acknowledgment} 

This work was partially supported by a grant
(No. SP/S2/013/93) by the Department of Science and Technology,
Government of India.

\newpage
\noindent{\bf Figure Captions}

\noindent Fig.1: The expected and observed number of Ly $\alpha$ lines
near the QSOs as a function of relative velocity w.r.t. the QSOs.  The
histogramme shows the observed number. Long dashed dotted line is for
HM96, solid line is for scaled HM96 , dotted line is for power law
background and dashed line is for pure galactic background assuming
single power law column density distribution. Long and short dashed
line is for scaled HM96, long dashed line is for pure galactic
background and dash dotted line is for power law background assuming
double power law column density distribution.\\

\noindent  Fig.2: $\chi^2$ probability as a function of the background
flux for higher systemic redshifts of the QSOs. The curves from right
to left are for systemic redshift higher by 0, 500, 1000, 1500 and 2000
km s$^{-1}$.\\
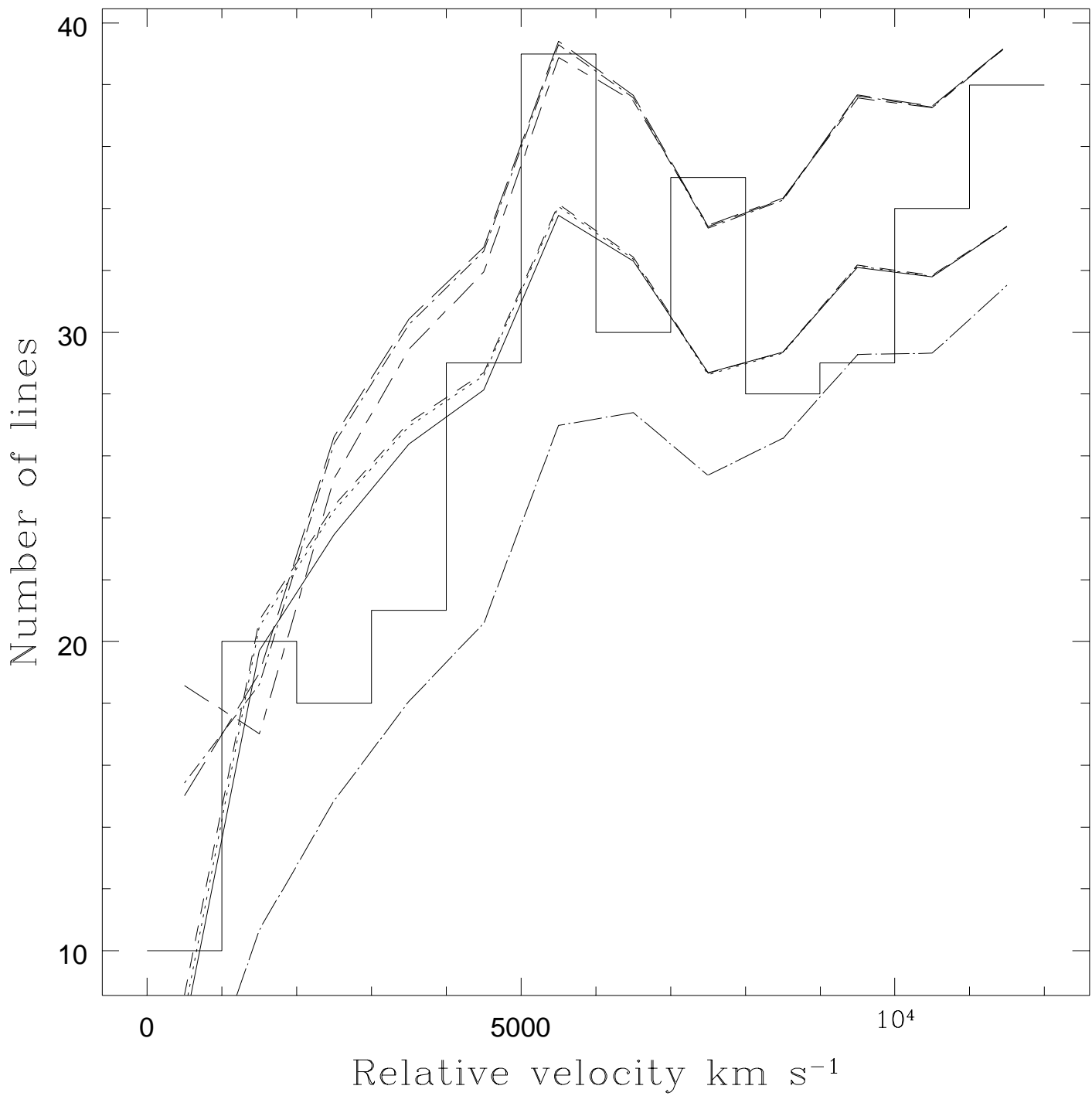
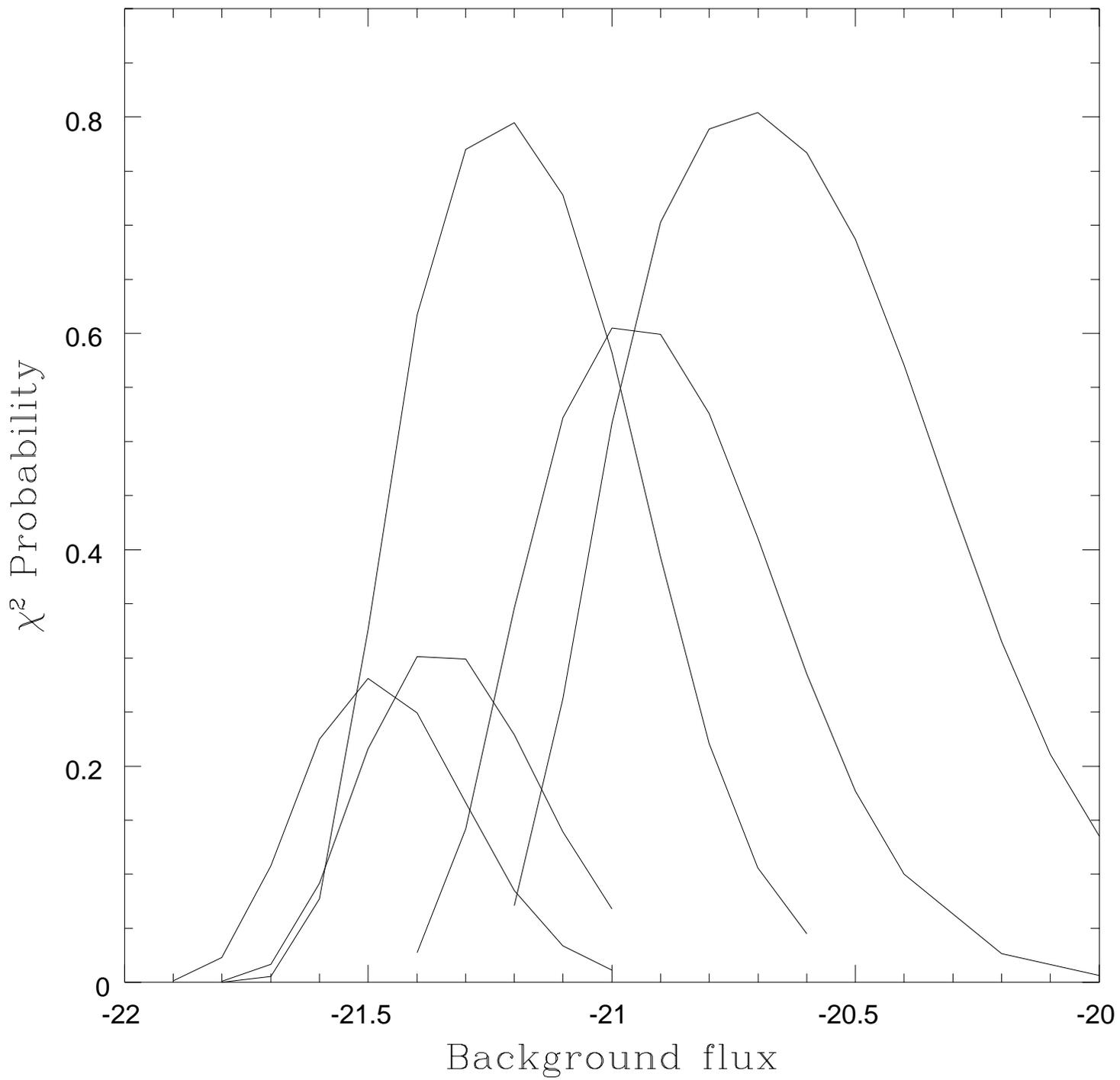
\end{document}